# National Research Impact Indicators from Mendeley Readers[1]

Mike Thelwall
Statistical Cybermetrics Research Group, School of Mathematics and Computer Science, University of Wolverhampton, Wulfruna Street, Wolverhampton WV1 1LY, UK.
Tel. +44 1902 321470. Fax +44 1902 321478, Email: m.thelwall@wlv.ac.uk
Ruth Fairclough
Statistical Cybermetrics Research Group, School of Mathematics and Computer Science, University of Wolverhampton, Wulfruna Street, Wolverhampton WV1 1LY, UK.
Tel. +44 1902 321000. Fax +44 1902 321478, Email: r.fairclough@wlv.ac.uk

National research impact indicators derived from citation counts are used by governments to help assess their national research performance and to identify the effect of funding or policy changes. Citation counts lag research by several years, however, and so their information is somewhat out of date. Some of this lag can be avoided by using readership counts from the social reference sharing site Mendeley because these accumulate more quickly than citations. This article introduces a method to calculate national research impact indicators from Mendeley, using citation counts from older time periods to partially compensate for international biases in Mendeley readership. A refinement to accommodate recent national changes in Mendeley uptake makes little difference, despite being theoretically more accurate. The Mendeley patterns using the methods broadly reflect the results from similar calculations with citations and seem to reflect impact trends about a year earlier. Nevertheless, the reasons for the differences between the indicators from the two data sources are unclear.

## 1. Introduction

Governments spend large amounts of money on academic research. For example, the National Science Foundation (NSF) was allocated $7.3 billion for this in 2015 (Rogers, 2014). Although some research supports health and quality of life improvements, the main purpose of the funding is to help enhance national competitiveness, particularly in the long term. Governments periodically change the amount of funding and the way in which it is allocated. For example, the UK has replaced fixed block research grants for universities with a competitive process based upon peer review, the precise nature of which changes every few years (Wilsdon, J. et al., 2015). One way to evaluate the ongoing performance of a nation's research and the effect of any policy changes is to evaluate the scientific impact of its research publications. This is an indirect indicator of success from the perspective of government because it does not directly reflect societal impacts, although these may be derived later, but has the advantage that it is straightforward to estimate in a relatively objective manner and the results can be compared over time to reveal trends in performance. The standard indicator for research impact is field normalised citation counts, for example by dividing the mean citation count of a set of articles by the world average for the same field. These allow international comparisons since these figures can be produced for the researchers of any country. For example, a report commissioned by the UK's Department for Business, Innovation & Skills included a graph comparing the relative

---

[1] Fairclough, R. & Thelwall, M. (2015). National research impact indicators from Mendeley readers. Journal of Informetrics, 9(4), 845–859. doi:10.1016/j.joi.2015.08.003.

citation impact of UK publications to those of eight other countries and the world average annually from 2008 to 2012 (Elsevier, 2013, p. 40). A similar approach is used by many other organisations and countries and for other purposes (e.g., Leydesdorff, Radicchi, Bornmann, Castellano, & Nooy, 2013; Science-Metrix, 2015; Waltman et al., 2012; Waltman, & van Eck, 2013; Zitt, Ramanana-Rahary, & Bassecoulard, 2005).

Whilst citation counts are an accepted indicator of academic impact, especially in the health and natural sciences, citations take time to accrue because of the delay between researchers reading an article and incorporating it into their research, as well as publication and peer review delays. Thus, whilst it would be possible to conduct citation analyses of recently-published articles, the citation counts would be low and even zero for most articles for a very recent analysis. This would make any indicators calculated from the citations more susceptible to outliers, such as sets of articles attracting particularly rapid citations due to being part of a themed special issue with editorial cross-citations, as well as individual articles attracting rapid citations due to being published in Early View by the publisher or self-archived by the author. Perhaps for this reason, early citations are not good predictors of long term citations (Levitt & Thelwall, 2011; Wang, 2013), although after a year the prediction may be reasonable (Adams, 2005).

Researching, peer review and publishing delays do not apply to some alternative indicators, such as counts of tweets, readers, or blog posts about articles. It therefore seems possible, in theory, to use these to develop field-normalised national impact indicators that show trends in advance of those in citation-based indicators, in the sense of more quickly accumulating enough data to be statistically robust. In practice, however, an alternative indicator must also reflect a type of impact reasonably well in order to give meaningful data. From this perspective, the most promising alternative to citation counts is readership counts from the social reference sharing site Mendeley because these have a moderate or high correlation with citation counts (Li, Thelwall, & Giustini, 2012; Li & Thelwall, 2012; Maflahi & Thelwall, in press; Thelwall & Sud, in press), in comparison to all of the alternatives (Haustein, Larivière, Thelwall, Amyot, & Peters, 2014; Thelwall, Haustein, Larivière, & Sugimoto, 2013) and also occur about a year earlier, on average (Maflahi & Thelwall, in press; Thelwall & Sud, in press). Although this makes the case for the value Mendeley for national research impact indicators, a method of constructing them is needed as well as evidence that the results are at least plausible. In response, this article introduces a technique to calculate Mendeley-based national research impact indicators as well as a second method that corrects for national biases in Mendeley uptake. These methods are assessed with an analysis of nine countries over 26 academic fields from 2009 to 2015.

## 2. Literature review

The research impact of a country's science base can be compared to the impacts of other countries using the average impact per paper, with field normalisation correcting for differing levels of citation (e.g., Elsevier, 2013, p. 40). This has the advantage of being size-independent so that it is easy to compare between countries. Nevertheless, productivity is important for a nation's science and so this information should be presented in conjunction with information about total publication counts (e.g., Albarrán, Crespo, Ortuño, & Ruiz-Castillo, 2010), perhaps normalised by the population, GDP or number of active researchers in each country.

When using citation counts for country comparisons, it is important to use fractional author citation counting (i.e., dividing the citation counts of each contributor by the number of contributors) rather than full citation counting, and to fractionalise based either on the number of affiliations or the number of authors (Aksnes, Schneider, & Gunnarsson, 2012; Huang, Lin, & Chen, 2011; Waltman & van Eck, 2015; Zheng, Zhao, Zhang, Huang, & Chen, 2014). Of these, fractionalising based on the number of authors seems intuitively to be a better approach because it allocates an equal share to each author. More complex approaches that allocate a greater share of credit to the first author, who tends to have made the greatest contribution (within science: Marusic, Bosnjak, & Jeroncic, 2011), are also possible but there is no agreed method for this and some disciplines use alphabetical authorship order instead (Engers, Gans, Grant, & King, 1999; Levitt & Thelwall, 2015). Hence the simple approach of sharing credit equally amongst authors, irrespective of order seems reasonable although it may be unfair to countries that tend to have first authorships in high impact international collaborative scientific papers.

Another problem is that citation databases have different levels of coverage of the academic outputs of nations. In particular although Scopus seems to be more comprehensive than WoS (Moed & Visser, 2008), WoS and Scopus both seem to have lower coverage of languages other than English (Aghaei Chadegani, Salehi, Yunus, Farhadi, Fooladi, Farhadi, & Ale Ebrahim, 2013; Archambault, Vignola-Gagne, Côté, Larivière, & Gingras, 2006; Li, Qiao, Li, & Jin, 2014; see also: Albarillo, 2014). This can also result in lower citation counts to non-English publications (Van Leeuwen, Moed, Tijssen, Visser, & Van Raan, 2001; Van Leeuwen, Moed, Tijssen, Visser, & Van Raan, 2011) and so will affect citation impact indicators as well as productivity indicators, particularly in the social sciences and humanities.

Citation counts also have many theoretical limitations for research evaluation purposes. Although citations within science can be created to acknowledge important prior work (Merton, 1973), they can also be created for negative reasons and may be influenced by irrelevant factors (Chubin & Moitra, 1975; MacRoberts & MacRoberts, 1989; Oppenheim, & Renn, 1978). Nevertheless, when compared between large enough numbers of publications, non-scientific reasons for citations tend to cancel each other out so that the resulting citation counts are reasonable indicators (but not measures) of overall scientific impact (van Raan, 1998).

## 2.1 Mendeley reader counts

The limitations of traditional citations, such as their reflection of scholarly impact rather than other types of impact, have led to the pursuit of alternative indicators for the impacts of academic outputs. These have included downloads to reflect usage (Moed, 2005), patent citations to reflect commercial impacts (Meyer, 2000; Roach & Cohen, 2013; Trajtenberg, 1990), syllabus mentions for educational impacts (Kousha & Thelwall, in press), clinical guideline mentions to reflect health benefits (Thelwall & Maflahi, in press), and web links or mentions to reflect general impacts (Cronin, Snyder, Rosenbaum, Martinson, & Callahan, 1998; Ingwersen, 1998; Vaughan & Shaw, 2003).

Academic research is sometimes mentioned by the users of large social web sites. For example, articles may be discussed in Facebook, posted about in Twitter or shared in social bookmarking sites. The combination of the large number of users of these sites and the relative ease with which their data can be accessed, often automatically through Applications Programming Interfaces (APIs), has led to the creation of many new alternative indicators. These fall under the umbrella term of altmetrics and, in theory, can usefully reflect wider impacts of research, such as amongst the general population (Priem, 2014). In order to assess the value of these new indicators, a logical first step is to assess whether they correlate with citation counts (Sud & Thelwall, 2014). Although positive correlations have been found for many altmetrics (Costas, Zahedi, & Wouters, 2014; Eysenbach, 2011; Shema, Bar-Ilan, & Thelwall, 2014; Thelwall, Haustein, Larivière, & Sugimoto, 2013), the correlations are highest for Mendeley readership counts and the coverage of most other altmetrics, apart from Tweet counts, is low (Zahedi, Costas, & Wouters, 2014). All altmetrics are also susceptible to spam (Wouters & Costas, 2012) and perhaps particularly Twitter (Haustein, Bowman, Holmberg, Tsou, Sugimoto, & Larivière, in press) due to its use for publicity by authors and publishers. Another Twitter drawback is that tweet counts seem to reflect casual interest rather than wider public engagement or scholarly impact (Thelwall, Tsou, Weingart, Holmberg, & Haustein, 2013). Nevertheless, as attention metrics, altmetrics can provide interesting metadata for journal readers (Adie & Roe, 2013), if not for evaluators.

There are several different major social reference sharing sites, including CiteULike, Zotero, Mendeley, and RefWorks. According to Alexa.com, the Mendeley.com attracts less users than do the websites of CiteULike and Zotero (global traffic ranks in July 2015: 14,322, 21,471, 55,306 and 24,532 respectively), although this data from Alexa's global user panel do not include uses of desktop applications. Each Mendeley member can add academic publications to their personal libraries in order to record them for future reference or to convert them into citations for a publication. Most articles in users' libraries have been read by them or are intended to be read by them (Mohammadi, Thelwall, & Kousha, in press) and so it is reasonable to interpret the Mendeley users of an article as readers of it. These users are predominantly academics, with a bias towards younger academics (Mas-Bleda, Thelwall, Kousha, & Aguillo, 2014; Mohammadi, Thelwall, Haustein, & Larivière, 2015), and so Mendeley readership counts for articles probably reflect the extent to which academics are interested enough to read them. The high correlations between Mendeley readership counts and citation counts suggest that it is also reasonable to interpret Mendeley readership as an indicator of academic impact (Li, Thelwall, & Giustini, 2012; Li & Thelwall, 2012; Maflahi & Thelwall, in press; Thelwall & Sud, in press). This conjecture is supported by the fact that both citation counts and Mendeley readership counts follow similar, but not identical, highly skewed statistical distributions (Thelwall & Wilson, in press).

Since Mendeley readership counts seem to reflect a similar type of impact to that of citations but can be spammed, they are not good candidates to replace or even complement citation counts for impact evaluation purposes. There are three exceptions to this: Mendeley readership counts are free and may be useful for those who cannot afford to access citation databases for scientometrics studies; occupation, nationality and subject area information is available for Mendeley readers and can be used to track knowledge flows between academic ranks, disciplines and countries (Mohammadi & Thelwall, 2014); and Mendeley readership counts accrue faster than do citations and so can be used for early impact evaluations. In terms of time, it seems that Mendeley readership counts accumulate about a year earlier than do citation counts (Maflahi & Thelwall, in press; Thelwall & Sud, in press). This is consistent with Mendeley users becoming citers as their articles get published, but it is likely that many or most Mendeley readerships do not directly translate into Scopus or WoS citations.

In addition to the bias of Mendeley towards younger users, its readership counts are likely to have national biases. This is because the uptake of Mendeley is not internationally uniform and the Mendeley readers of an article are disproportionately from the countries that the authors are affiliated with (Thelwall & Maflahi, 2015). Although this bias is not substantial, it is a problem when using Mendeley for international comparisons and so corrective steps must be taken.

# 3. Research questions

The evidence discussed above suggests that Mendeley-based indicators should lead citation-based indicators by about a year but that there may be national biases in uptake of Mendeley that could translate into national biases in readership counts. Hence, the logical primary way to assess readership-based indicators is by comparing them with citation-based indicators on the basis that they should be stable for more recent years. This is operationalised in the following research questions.
1. Can Mendeley-based national impact indicators give credible results for recent years?
2. Can Mendeley-based national impact indicators give more stable results than citation-based national indicators for recent years?

# 4. Data and Methods

## *4.1 Data*

Ideally, impact indicators should be calculated from a complete set of all of the academic outputs of all countries in order to give comprehensive results. In practice, however, data must be collected from an established database with reasonably good international coverage, even if it is not comprehensive. The two largest international databases of academic outputs are the Web of Science (WoS) and Scopus, with the latter being preferable for its wider international coverage (Erfanmanesh & Didegah, 2013; Li, Burnham, Lemley, & Britton, 2010; Minasny, Hartemink, McBratney, & Jang, 2013). Although academics produce a variety of outputs, including books and conference papers, Scopus has very limited and nationally biased coverage of books (Torres-Salinas, Robinson-García, Campanario, & López-Cózar, 2014) and so these should not be included. Scopus probably has wider coverage of conference proceedings and these have value comparable to that of journal articles in some fields, such as computer science (Goodrum, McCain, Lawrence, & Giles, 2001) and are important in some areas of engineering and technology (Lisée, Larivière, & Archambault, 2008). Nevertheless, conference papers are not valued in other areas and so it is better to exclude them as a universal rule. Hence, the data set was restricted to journal articles from Scopus. To further ensure uniformity, documents within journals that were marked in Scopus as anything other than an article (e.g., review, editorial) were excluded.

In order to ensure widespread coverage of academic research, 26 different Scopus categories were selected for the indicators. These were chosen to represent most areas of research (Animal Science and Zoology; Language and Linguistics; Biochemistry; Business and International Management; Catalysis; Electrochemistry; Computational Theory and Mathematics; Management Science and Operations Research; Computers in Earth Sciences; Finance; Fuel Technology; Automotive Engineering; Ecology; Immunology; Ceramics and Composites; Analysis; Anesthesiology and Pain Medicine; Biological Psychiatry; Assessment and Diagnosis; Pharmaceutical Science; Astronomy and Astrophysics; Clinical Psychology; Development; Food Animals; Orthodontics; Complementary and Manual Therapy). The indicators were normalised separately for each of these fields, assuming that each field formed a relatively coherent collection. Although the Scopus categories are imperfect and alternative methods of defining fields are available, such as through citation, reference or keyword clustering (Braam, Moed, & van Raan, 1991), and other methods of normalisation have been developed, such as based on the number of references in the citing source (Waltman & van Eck, 2013; Zitt & Small, 2008), the use of Scopus categories has the advantages of being transparent, simple and reasonably intuitive.

Information about all articles in each of the chosen categories published between 2009 and 2015 was downloaded from Scopus between April 15 and May 11, 2015, together with their Scopus citation counts. For subject categories with more than 10,000 documents in a single year, only the first 5,000 and last 5,000 documents could be extracted and so the data sets are incomplete for these years but this should not systematically bias the results. Authorship information for each article, including the country of their main affiliation was extracted from Scopus. Authors without an assigned country within Scopus were kept in the data set and allocated the dummy country affiliation, "NA" (6%) so that they would not be excluded from the calculations.

Mendeley readership counts were extracted for each article May 7-17, 2015 using Webometric Analyst (http://lexiurl.wlv.ac.uk). For each article in the Scopus data set the Mendeley Applications Programming Interface (API) was queried for the title, year and first author of the article and the list of matches recorded. When present in Scopus, the article Digital Object Identifier (DOI) was also queried for additional matches (Zahedi, Haustein, & Bowman, 2014). The readership counts were combined for all correct matches found, after eliminating obvious incorrect matches such as those with an incorrect year or DOI (for more details see: Thelwall & Sud, in press).

## *4.2 Field and year normalised citation and readership indicators*

Each article was assigned to countries in accordance with the country's share of authorship, making the simplifying assumption that all authors had contributed equally. Hence, an article with three Spanish and one Argentinian author would count as 0.75 Spanish and 0.25 Argentinian. The proportion $p_{c,a}$ contribution of nation state (i.e., country) $s$ to article $a$ with $n_{s,a}$ authors from nation state $s$ and $n_a$ authors in total is: $p_{s,a} = n_{s,a}/n_a$. Let $A_y$ denote the set of all articles in year $y$. Then the fractional author citation counting publication output of country $s$, which is the sum of the contributions to each article, is given by: $o_{s,y} = \sum_{a \in A_y} p_{s,a}$.

Let $A_{f,y}$ be the set of articles in field $f$ and year $y$. Then the total number of articles in a field and year is $|A_{f,y}|$, the size of the set $A_{f,y}$. Setting $c_a$ to be the number of citations to article $a$, the total number of citations attracted by articles in field $f$ and year $y$ is $\sum_{a \in A_{f,y}} c_a$ and so the average number of citations per article is this total divided by the number of articles: $\sum_{a \in A_{f,y}} c_a / |A_{f,y}|$. The field normalised citation count for each article is then obtained by dividing its citation count, $c_a$, by the above citation average. For article $a$ in field $f$ and year $y$, its *field and year normalised citation count* is therefore $FYNCC_a = c_a/(\sum_{a \in A_{f,y}} c_a / |A_{f,y}|)$.

Similarly, setting $r_a$ to be the number of Mendeley readers of article $a$, the total number of readers attracted by articles in field $f$ and year $y$ is $\sum_{a \in A_{f,y}} r_a$ and so the average number of readers per article is $\sum_{a \in A_{f,y}} r_a / |A_{f,y}|$. For article $a$ in field $f$ and year $y$, its *field and year normalised readership count* is therefore $FYNRC_a = r_a/(\sum_{a \in A_{f,y}} r_a / |A_{f,y}|)$.

For each year, the *average field and year normalised citation count of country s* in year *y* is the total number of field normalised citations of its articles divided by its number of articles, taking into account authorship shares in both cases.

$$AFYNCC_{s,y} = \left(\sum_{a \in A_y} FYNCC_a \times p_{s,a}\right)/o_{s,y} \tag{1}$$

The readership calculations follow the same pattern, so the *average field and year normalised readership count of country s* in year *y* is given by a similar formula.

$$AFYNRC_{s,y} = \left(\sum_{a \in A_y} FYNRC_a \times p_{s,a}\right)/o_{s,y} \tag{2}$$

## *4.3 Field and year normalised, national bias corrected readership indicator*

A national bias correction figure is needed to compensate for Mendeley being used less in some countries than in others and users tending to read articles from their own country more than articles from other countries (Thelwall & Maflahi, 2015). A logical way to identify such a correction factor would be to obtain data on the national spread of Mendeley users or the national spread of Mendeley readers. The latter could be estimated, for example, from the Mendeley readership data collected for the 26 fields analysed here. This is not enough, however, because an estimate would also be needed for each country of the bias that they show towards their own articles. A simpler approach would be to assume that old articles had received most of their citations and readers and so any time lag factors between readers and citers would be small. If this assumption is true then the difference between the normalised citation impact and the normalised readership count would be a good estimator of the overall bias in the system. The time needed to attract nearly all citations is too long to be practical however, since the median age of the cited literature in natural sciences and engineering articles (cited half-life) was about 7 in 2004 and for medicine the median age was 5.5 (Larivière, Archambault, & Gingras, 2008; see also: Davis, & Cochran, 2015; Glänzel & Schoepflin, 1999). Nevertheless, correcting with the oldest possible data should reduce the extent of the Mendeley readership bias. The use of a more recent year risks cancelling out the lead of readership counts over citations but a much older year may give incorrect bias corrections due to changes in international patterns of Mendeley use, assuming that users tend to register recent articles in Mendeley. In practice, however, the year used can be about five years in the past because the correlation between citations and readership counts is approximately constant after five years (Thelwall & Sud, in press), which suggests that older years may not give an advantage.

Using the above reasoning, the (under-)estimated amount of readership bias in the system for country *s* is $b_s = AFYNRC_{s,y_0}/AFYNCC_{s,y_0}$, where $y_0$ is about five years in the past. This assumes that the bias is constant between fields and does not change over time, which are oversimplifications. A bias value of greater than 1 for a country suggests that its normalised readership counts overestimate its normalised citation counts, perhaps because it has many Mendeley users.

The *average citation-corrected field and year normalised readership count of country s* in year *y* is therefore obtained by dividing by the bias correction factor.

$$ACCFYNRC_{s,y} = AFYNRC_{s,y}/b_s \tag{3}$$

This bias correction should incorporate all sources of bias from Mendeley, such as due to uptake and differing uses of DOIs, but not biases that affect Mendeley and Scopus citations in the same way. Although the world average for formula (1) and (2) is 1, the introduction of a bias correction factor may change the world average for formula (3) to be higher than 1 if the readership values $AFYNRC_{s,y}$ tend to be above 1 for countries with a bias correction factor $b_s > 1$. The opposite can also occur and so the world average for this statistic is not necessarily 1, although it should be close to 1. The results could be corrected to make the world average 1 but this is not done here.

The above calculations do not use a citation window but include all citations to date. The readership calculations also include all readerships to date. This approach uses the most data available and the comparisons between years are based upon normalised values. Nevertheless, some of the differences between years may be due to shortening citation windows since some countries tend to attract citations more quickly than do others, even within the same field (Jonathan Adams, personal communication).

## 5. Results

The three impact statistics were calculated as above except that the citation/readership correction factor was calculated from the average of 2009 and 2010. This seemed to be more reliable than using either the 2009 or the 2010 value because the basic readership indicators changed from 2009 to 2010 by a relatively large amount for Italy (Figure 2), and so averaging 2009 and 2010 is a conservative step to reduce the possibility that 2009 was an unusual year in some way.

Figure 1 mimics the methods of a previous study (Elsevier, 2013) that used all Scopus articles (2008-2012) rather than a subset of 26 categories, whole author counting rather than fractional counting, and citation data from 2013 rather than from 2015. Figure 1 gives broadly similar results for the sets of years that they both cover (2009-2012), but with some substantial differences. The Russian Federation increased from 0.45 to 0.6 in Figure 4.6 of the prior report (Elsevier, 2013) but is much lower at a constant 0.3 in Figure 1. The reason for the difference is that the Russian Federation must have a higher average normalised citation rate across Scopus in comparison to the 26 subjects selected for Figure 1. For example, it is relatively successful in physics and space sciences (Adams & King, 2010), and only Astronomy and Astrophysics was included in Figure 1. Similarly, the UK, USA and Canada are between 1.4 and 1.6 in Figure 4.6 (Elsevier, 2013) but between 1.2 and 1.4 in Figure 1. These differences are probably due to underrepresenting areas of particular strengths of these countries (e.g., health sciences and/or natural sciences) in the sample of 26 subjects. The choice of a wide range of different subjects does not give a representative sample of Scopus because Elsevier indexes far more articles in some broad categories than others (e.g., there are 48 subcategories of the broad category Medicine but only three subcategories of the broad category Economics, Econometrics and Finance). There are also small differences in rankings due to the coverage issue. For example, the UK is above the USA throughout Figure 4.6 (Elsevier, 2013) but below during the overlapping period (2009-2012) in Figure 1. This is probably due to the UK having a particularly large citation impact in areas (e.g., health) that are particularly underrepresented in Figure 1 compared to Scopus. The use of whole author counting rather than fractional counting in the previous study will also affect the results for countries with many internationally co-authored papers.

Another important source of differences between the results is the longer citation window used in Figure 1. If some countries tend to attract citations more rapidly than do others in the same specialty then shorter time periods would give them an advantage for international comparisons (Jonathan Adams, personal communication). This seems likely to give a small advantage to countries with English as a first language because this can help speed the production of their articles since they do not have to wait for one or more rounds of language polishing, may write-up their research more quickly in the first place and in some countries and disciplines may also write journal articles in their native language, with these being less likely to be indexed by Scopus or WoS. Hence their references might be slightly more recent, and authors tend to cite other researchers from the same country disproportionately often. It is not clear how big this advantage is, however.

After 2012, the lines in Figure 1 are relatively smooth except for 2015, giving credibility to the results up to 2014. The Canada result for 2015 must be an anomaly due to the low data coverage since it reverses the previous slowly decreasing trend to an extent that seems highly unlikely.

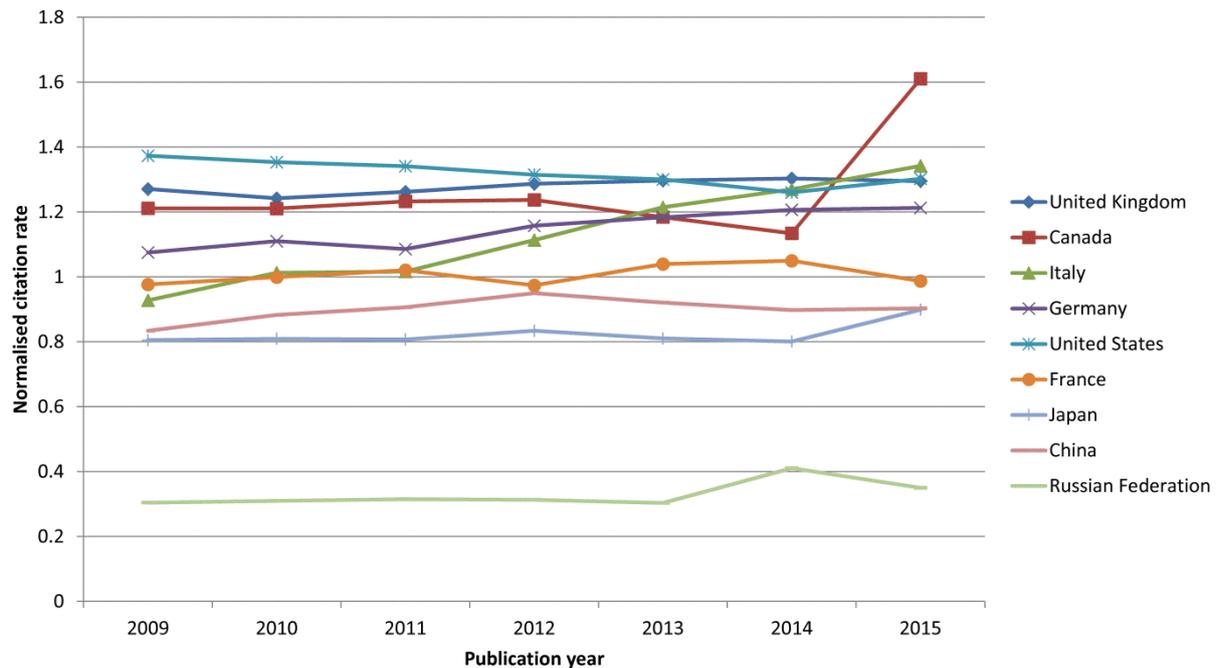

Figure 1. Average field and year normalised citation count $AFYNCC_{s,y}$ for 26 Scopus subjects by authorship country (fractional counting) and year (n=923,711 articles). The world average is 1.

The results from Mendeley (Figure 2) are only very broadly similar to the Scopus citation results (Figure 1) in terms of overall patterns and country rankings and there are substantial differences. The UK is dominant in Figure 2 from 2012 but not in Figure 1, for example, and the values tend to be further away from the overall average rate of 1. The lines are also a little smoother for more recent years, suggesting that the hypothesised early data benefit of Mendeley has made the results more powerful.

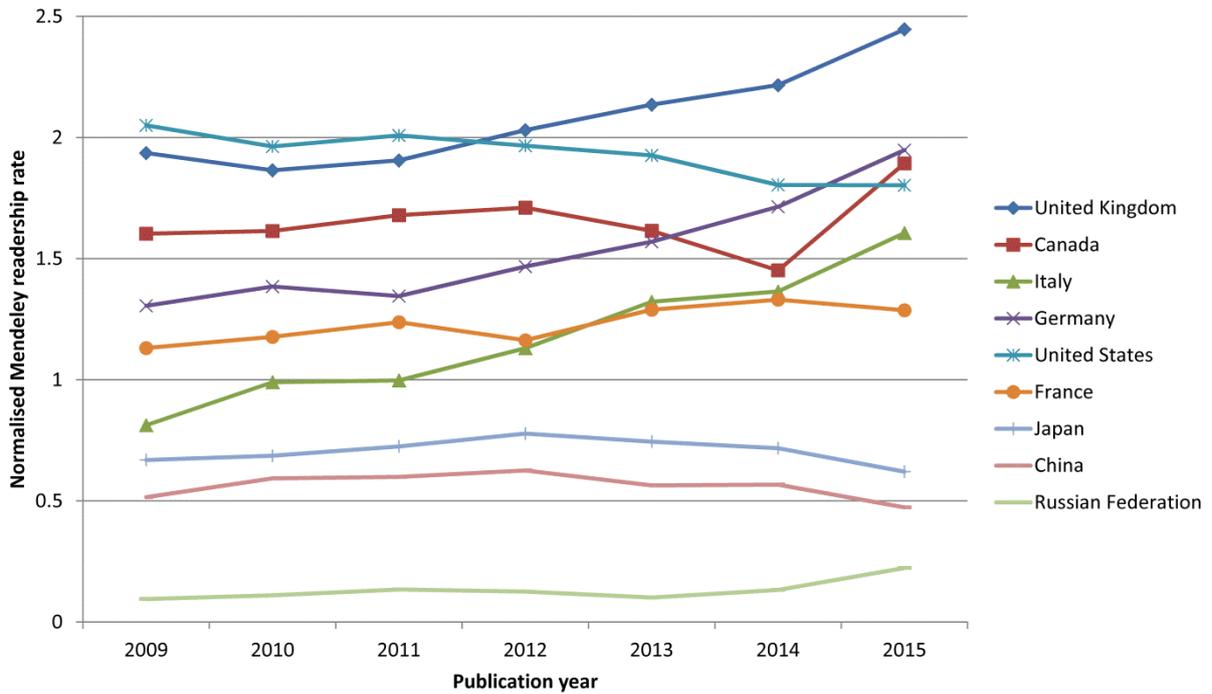

Figure 2. Average field and year normalised readership count $AFYNRC_{s,y}$ for each of the 26 Scopus subjects by authorship country (fractional counting) and year (n=923,711 articles altogether). The world average is 1.

The citation-corrected Mendeley readership rates graph (Figure 3) should give better results because it corrects for national biases in the uptake of Mendeley and other national biases in Mendeley readership (see Figure 4). In answer to the first research question, the results look consistent, in the sense of being relatively smooth and following a pattern, until 2014 but not for 2015 because of the Canada anomaly. The results also seem reasonable in the sense that the overall rankings are not very different from those in Figure 2, although there are some differences. Hence, the results appear visually to be broadly credible.

The graph (Figure 3) suggests for the first time that Italy has overtaken the UK and the USA for average citation counts, but this may be an artefact of the selection of 26 subjects covered. Nevertheless, the rapid and sustained growths of Italy, Germany and the UK, at least in these 26 subjects, are impressive.

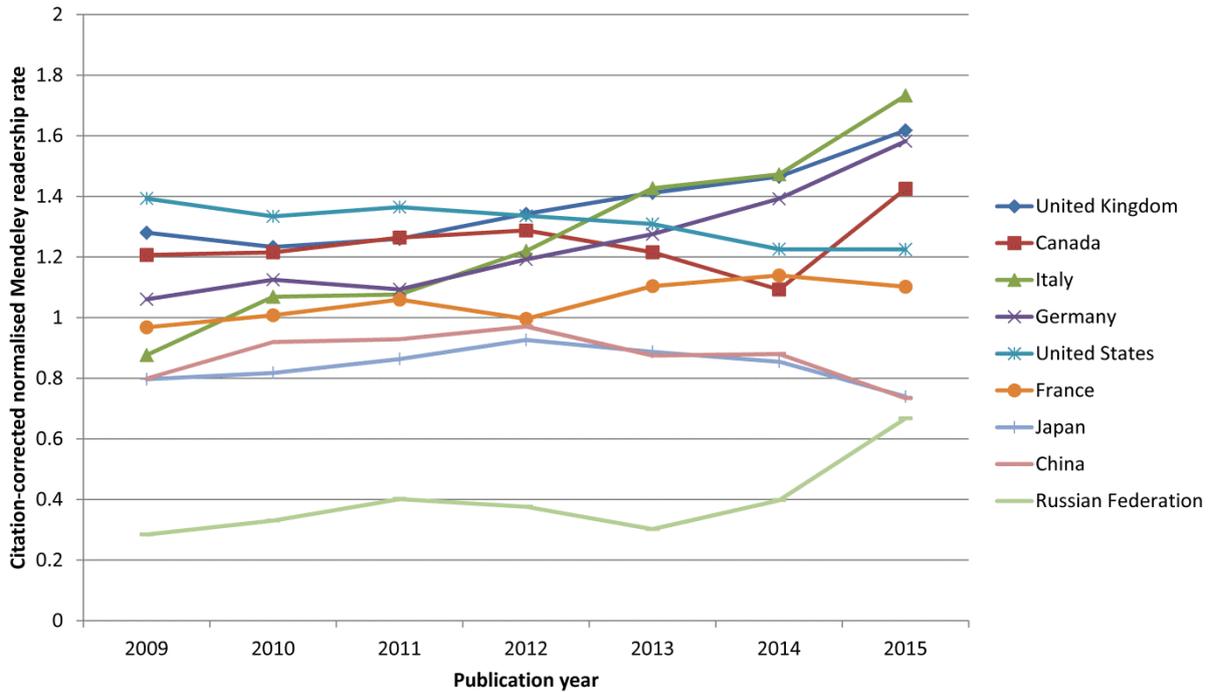

Figure 3. Average citation-corrected field and year normalised readership count $ACCFYNRC_{s,y}$ for 26 Scopus subjects by authorship country (fractional counting) and year (n=923,711 articles altogether). The world average is approximately 1.

Surprisingly, the trend in Figure 3 seems to lead the trend in Figure 1. For example, the UK overtakes the US in 2012 in Figure 3 but in 2013 in Figure 1. Similarly the Italy overtakes both the UK and USA in Figure 3 earlier than in Figure 1.

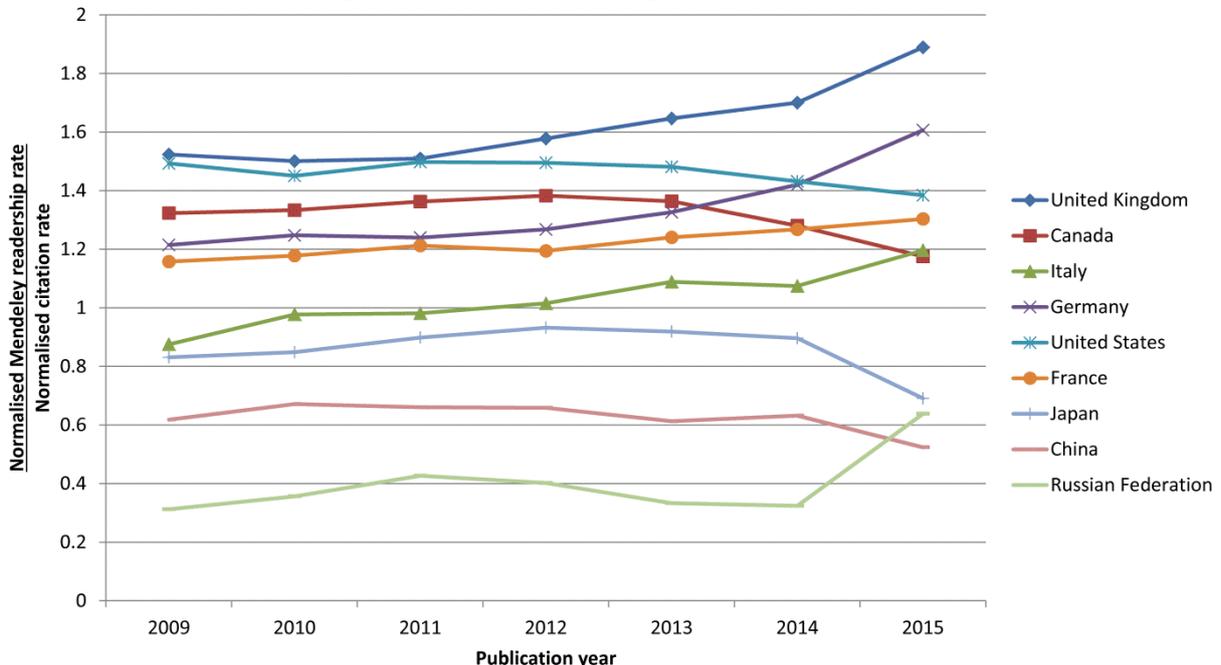

Figure 4. Average field and year normalised readership count for 26 Scopus subjects by authorship country (fractional counting) divided by the average field and year normalised citation count for 26 Scopus subjects by authorship country (fractional counting) $AFYNRC_{s,y}/AFYNCC_{s,y}$ (n=923,711 articles altogether).

There are four times as many readers as citations in 2015 and almost four times as many in 2014 (Table 1). The magnitude of these differences means that statistics with readership data for these years will be much more powerful than statistics with citation data. This might not matter for normally distributed data because of the large numbers involved but for highly skewed citation (and readership) data, the extra power seems to be important. Hence, other factors being equal, Mendeley readership data would be preferable for the current year and the previous year for country-level research impact comparisons, answering the second research question.

Table 1. Summary statistics for the data collected.

| Year | Articles | Citations | Readers | Citations per article | Readers per article | Citations per reader |
| --- | --- | --- | --- | --- | --- | --- |
| 2009 | 137905 | 1975043 | 1625447 | 14.32 | 11.79 | 1.22 |
| 2010 | 139125 | 1635585 | 1654037 | 11.76 | 11.89 | 0.99 |
| 2011 | 141600 | 1227703 | 1500801 | 8.67 | 10.60 | 0.82 |
| 2012 | 146359 | 863505 | 1338807 | 5.90 | 9.15 | 0.64 |
| 2013 | 151069 | 489583 | 1047641 | 3.24 | 6.93 | 0.47 |
| 2014 | 136137 | 128907 | 477214 | 0.95 | 3.51 | 0.27 |
| 2015 | 71516 | 8426 | 36416 | 0.12 | 0.51 | 0.23 |
| Total | 923711 | 6328752 | 7680364 | 6.85 | 8.31 | 0.82 |

# 6. Discussion

The differences between Figure 1 above and Figure 4.6 from the full Scopus database (Elsevier, 2013) highlight the importance of the coverage of a citation database and perhaps also the citation window for the findings of an international comparison exercise. The coverage of the citation database can affect the results in relatively obvious direct ways as well as in indirect ways. The direct influence is the extent of coverage of different subject areas. The results reflect the subject coverage of Scopus and so are skewed in favour of countries that perform well in the areas that Scopus over-represents. For example, if the Scopus coverage of computing is particularly comprehensive then countries that have computing strengths tend to be overrated. Given that international comparisons focus on refereed journal articles, countries with strengths in the arts and humanities are disadvantaged because these tend to produce books and non-documentary outputs. In addition, Scopus has more comprehensive coverage of English than of other languages and so countries with strong social sciences research that is frequently published in national literature are also disadvantaged. Although the coverage of Scopus (or WoS) has a clear and direct effect on the results, the nature of this effect is hard to gauge because there no effective way to assess the relative coverage of Scopus in any given academic field.

The coverage of Scopus also has an indirect effect on the results due to the normalisation process used. The results favour countries that tend to publish in high impact journals within each specialty because these tend to be cited above the world average for their specialty, as calculated by Scopus. The results are therefore biased against countries in which Scopus has particularly comprehensive coverage of the national literature especially if that national literature is weak. This has been shown to affect specific subject areas, such as oncology (López-Illescas, Moya-Anegón, & Moed, 2009), and may also affect national literatures more generally. It can also lead to spurious increases or decreases in countries over time. For example, if Elsevier expands its coverage of Chinese-language journals by indexing lower quality Chinese journals than previously then the average normalised citation score of China will decrease due to the influx of low impact articles and the average normalised citation score of all other countries will increase because the average citation score used for normalisation will decrease.

The five year span for the citation normalisation calculation for Mendeley is a problem because it assumes that national uptake has been approximately uniform for those five years. Moreover, a country could change its disciplinary specialty to some extent over five years and might as a result move into an area where it has particularly high or low Mendeley uptake and this would also affect the citation correction calculations. The national uptake of Mendeley in the relevant subject areas can be estimated by counting the declared national affiliations of the readers of these articles for the minority of readers with such a declared affiliation. The results suggest that there have been changes in the national shares of Mendeley readers, with Italy steadily increasing its share but the USA and Germany both reducing their share (Figure 5). In percentage change terms, Italy's proportion of readers increased by more than 20% of the initial value from 2009-2010 to 2015 (Figure 6). If this is due to an increased use of Mendeley by Italian researchers rather than due to increasing research or publishing by Italian scientists, then this would increase the bias in the use of Mendeley readers as a proxy for Scopus citations.

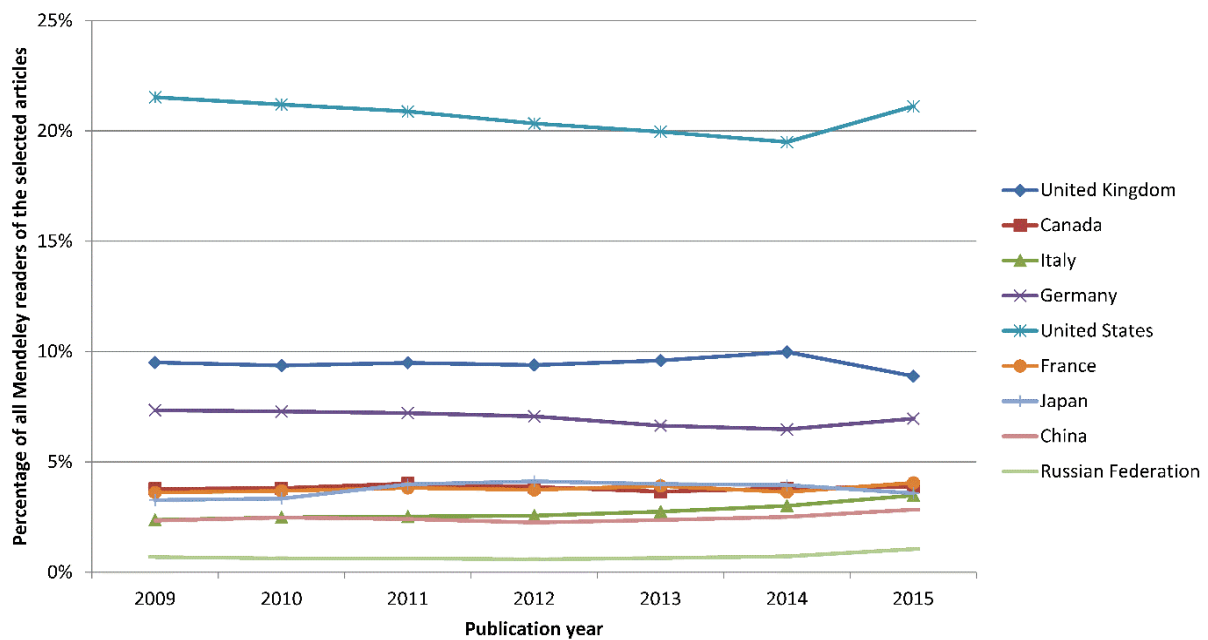

Figure 5. Percentage of Mendeley readers of the articles in 26 Scopus subjects by declared country of the readers, when declared (923,711 articles and 7,680,364 readers altogether).

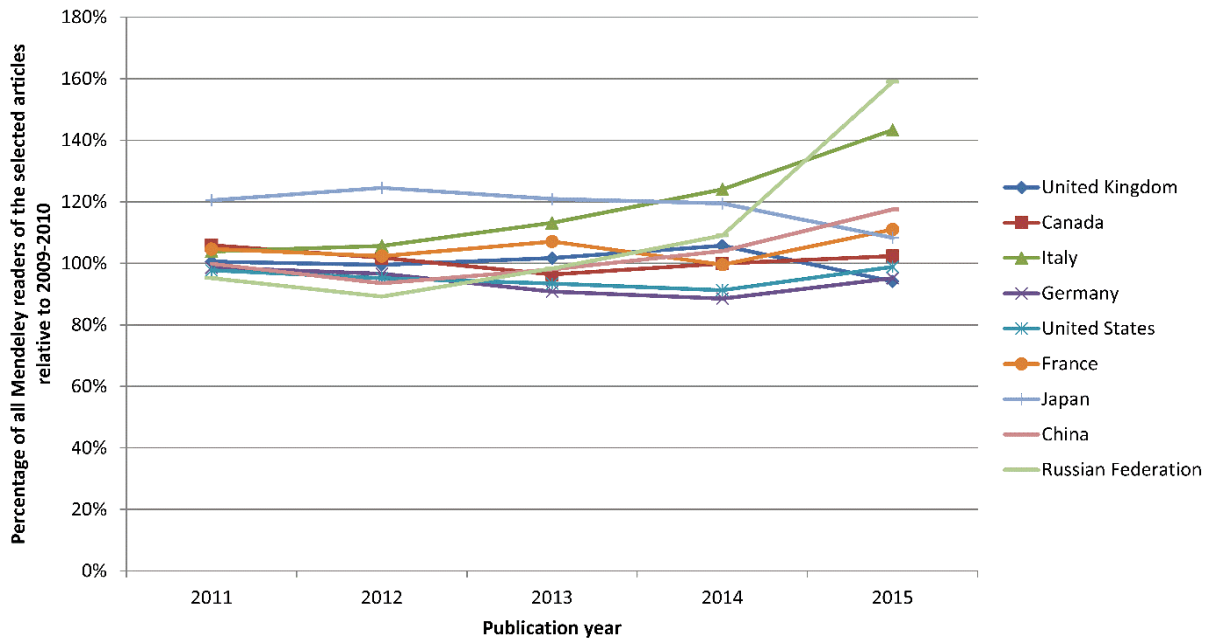

Figure 6. Percentage change in the number of Mendeley readers of the articles in 26 Scopus subjects by declared country of the readers, when declared (646,481 articles and 4,400,879 readers altogether, excluding 2009-2010).

The changing percentage of the national share of Mendeley readership (Figure 6) could be used to further correct the Mendeley-corrected estimated impact calculations (Figure 3). For this, however, an estimate is needed for the amount of national bias in the readership of each country. The data confirms that national biases exists in Mendeley readership and are substantial because the proportion of articles read by people from the same country as the author (Figure 7) is higher than the proportion of articles from that country by the rest of the world (Figure 8). Subtracting the two gives an estimate of the national bias for each country (Figure 9) in terms of the percentage of same country readerships that could be attributed to national bias. Some of this bias may be due to the normal tendency for countries to specialise so that they would naturally take more of an interest in national than international research, on average, and so the data in Figure 9 is likely to overestimate the amount of national bias that should be ignored in the calculations.

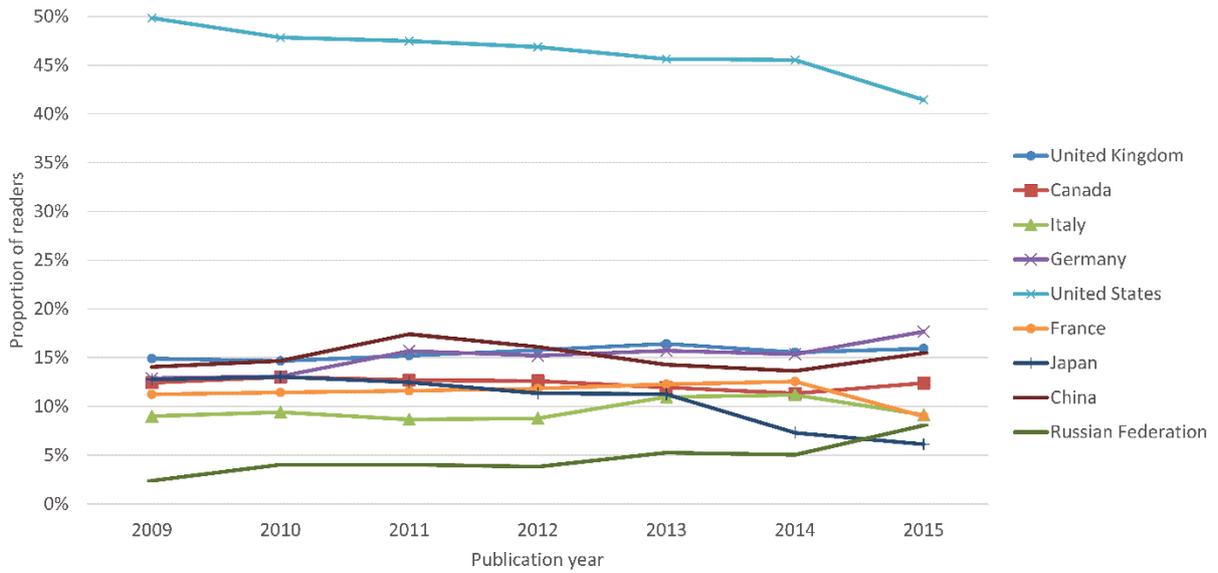

Figure 7. The proportion of Mendeley readers for articles authored within a country (fractional counting) and readers from the same country for 26 Scopus subjects.

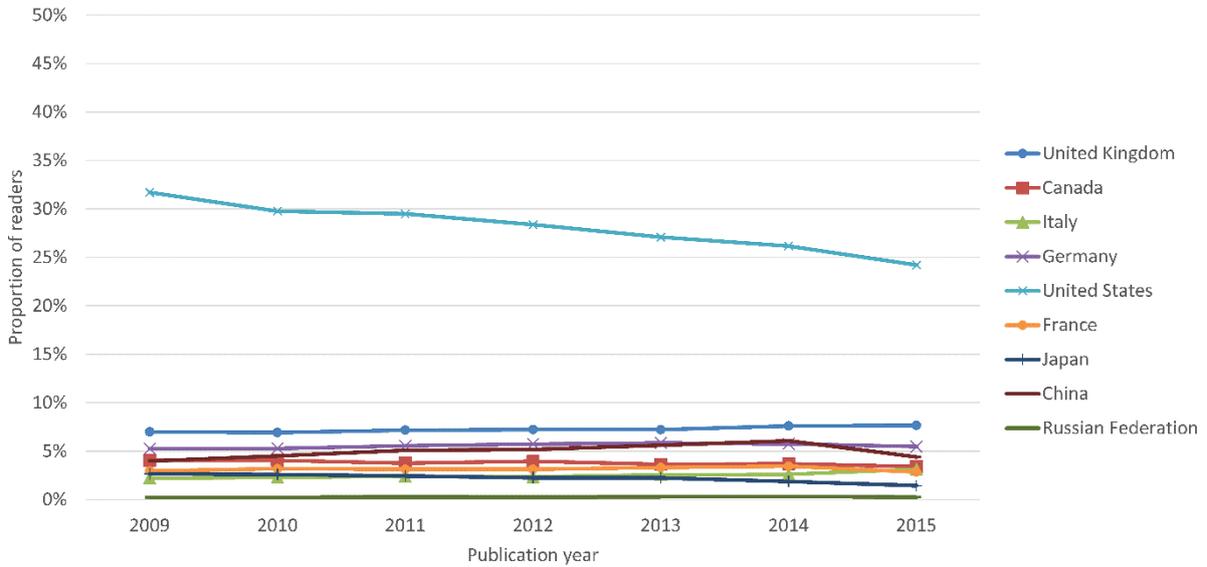

Figure 8. The proportion of Mendeley readers for articles authored within a country (fractional counting) and readers from a different country for 26 Scopus subjects.

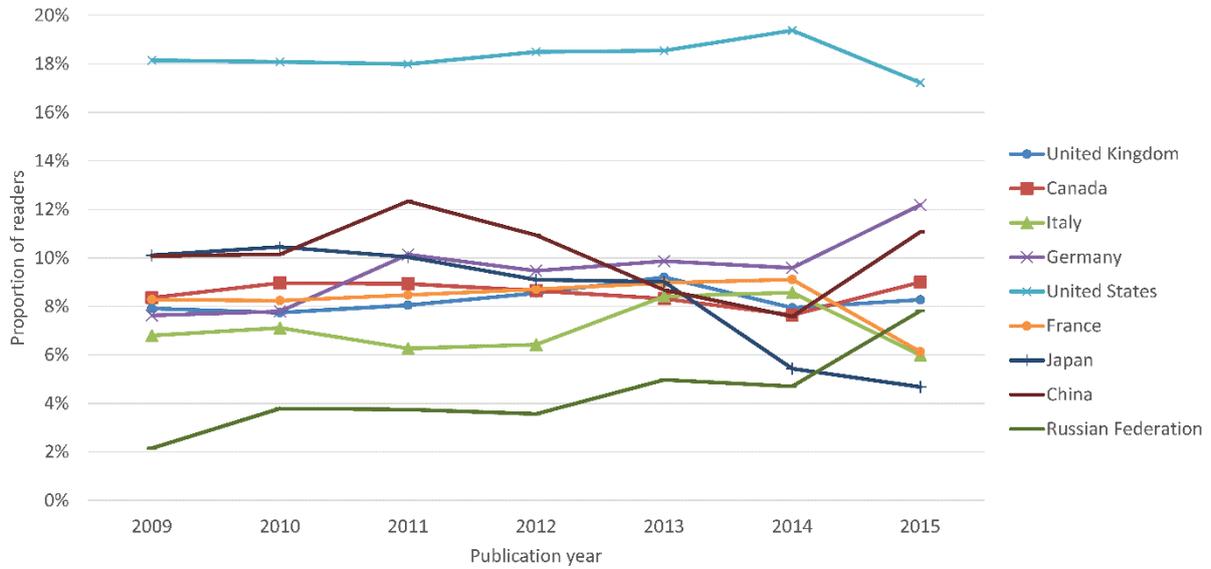

Figure 9. The national bias in reading habits expressed as the Figure 8 subtract Figure 7.

Doubly-corrected national impact comparisons can now be made to compensate the Figure 3 data for national changes in uptake by dividing the best previous indicator by the expected amount of increase that is due to the increase in the proportion of Mendeley readers from that country. This expected increase is the gross percentage change in readers from a country compared to the reference year, $incr_{s,y}$, times the bias figure (Figure 9) for that country, which is the rate at which it cites above the world average $self_{s,y}$. The *average citation-corrected field, year and Mendeley uptake change normalised readership count of country s* in year *y* is therefore obtained by dividing by the bias correction factor. This formula is an approximation because it does not take into account the effect (identical in the formula for all countries) due to the proportional reduction in the number of readers from the rest of the world due to an increase in the number of readers from the country *s* in the formula.

$$ACCFYMUCNRC_{s,y} = ACCFYNRC_{s,y}/(incr_{s,y} \times self_{s,y}) \qquad (4)$$

The results (Figure 10) are almost indistinguishable from formula (3) (Figure 3) because the correction factors are very small and all are less than half of one percent. The largest correction factor is for the United States in 2014, for which $1/(incr_{s,y} \times self_{s,y}) = 1.004$. This accounts for a 19% pro-US bias in readership for US readers in 2014, and a 2% decrease in readers from the US from 2009-10 to 2014 (i.e., of all Mendeley readers of articles, there were 2% less from the USA in 2014 in comparison to 2009-10). The reason why the corrections are so small is because in most cases the vast majority of the readers of a country's articles derive from other countries rather than the country in question and because they multiply two factors: small changes in gross percentages of Mendeley readers and country bias.

The calculations for Figure 10 also assume that the Mendeley bias is due only to the national uptake of Mendeley. This assumption is not true because some countries read articles from specific other countries more (or less) than average (Thelwall & Maflahi, 2015), and so the bias is partly due to the uptake by countries other than the one analysed.

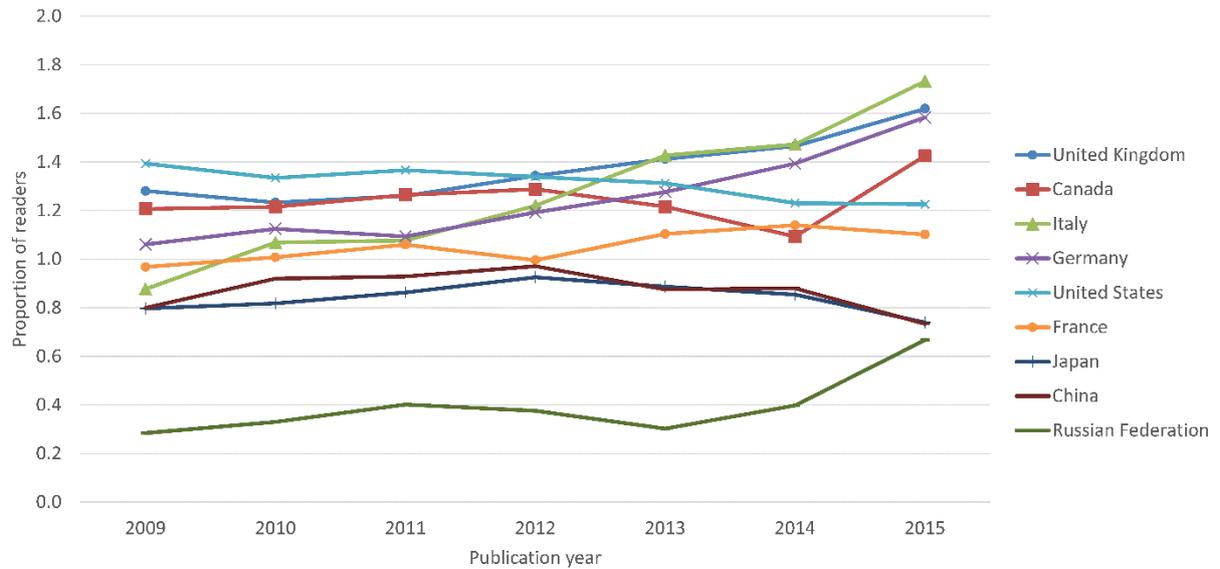

Figure 10. Mean citation-normalised, national Mendeley readership normalised, Mendeley readership count for 26 Scopus subjects by authorship country (fractional counting) and year, $ACCFYMUCNRC_{s,y}$. The world average is approximately 1.

A more direct approach would be to correct for the national biases in Mendeley uptake by calculating it directly for each year and country and then by dividing by this figure to cancel out this national bias. This is the gross percentage of readers from a country $share_{s,y}$ times the bias figure $self_{s,y}$ for that country (Figure 11). This assumes that national biases are only due to national uptake patterns, but this excludes the possibly more important factor of the extent to of Mendeley uptake in subject areas in which the country specialises. It also excludes any other sources of bias, such as younger and possibly more technological Mendeley users tending to favour one country over another. The *average field, year and Mendeley uptake change normalised readership count of country s* in year *y* is therefore obtained by dividing by the bias correction factor. This is a simplification since the calculation should also be adjusted to remove the "extra" readers for each country from the denominator of $AFYNRC_{s,y}$ but this omission is the same for all countries and so does not affect comparisons between countries.

$$AFYMUCNRC_{s,y} = AFYNRC_{s,y}/(share_{s,y} \times self_{s,y}) \qquad (5)$$

This bias correction makes all of the figures lower but makes little difference in terms of the relative positions between countries (compare Figure 11 to Figure 2) with the main visible change being for the increased closeness between the USA and UK during 2009-2011.

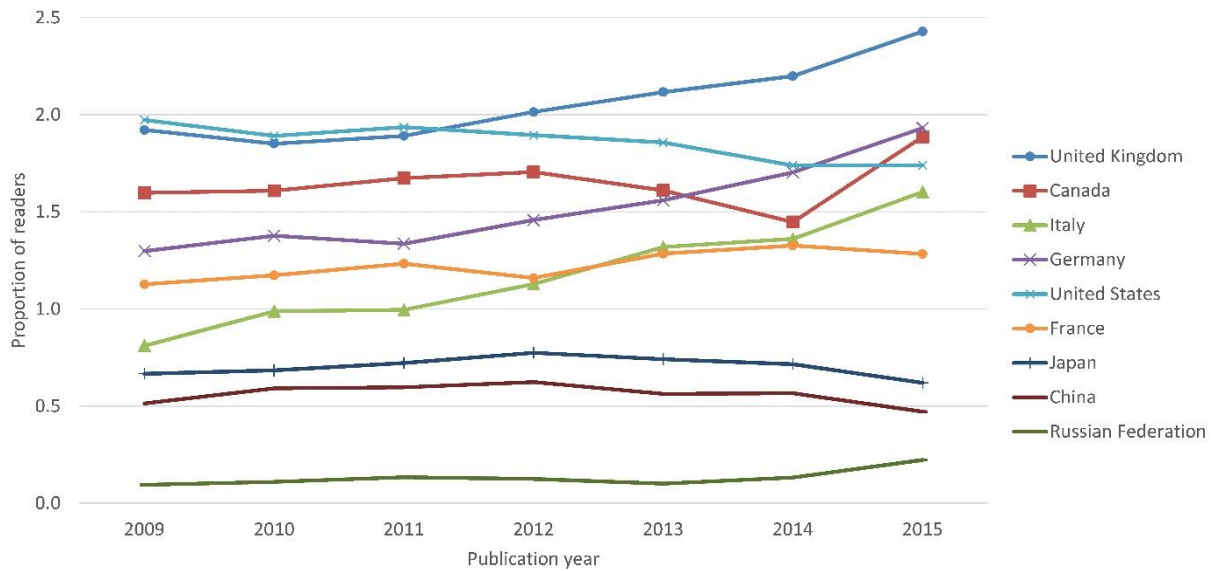

Figure 11. Average field and year normalised readership count $AFYMUCNRC_{s,y}$ for each of the 26 Scopus subjects by authorship country (fractional counting) and year, divided by the proportion of extra readers gained through apparent national bias. The world average is approximately 1.

## 7. Conclusions

The evidence shows that international research impact calculations using Mendeley data may be slightly more stable for recent years than are calculations based upon citation counts, although the difference is not large (Figure 1 vs. Figure 2) and so the evidence is weak. The slightly increased stability for the readership data is a logical expectation due to the much larger number of readers of than citations to articles from the past two years (Table 1). The stability appears to be enough to allow international comparisons for a publication year ending within five months of the data collection. The use of Mendeley readers introduces a bias due to differing national levels of uptake of Mendeley and a tendency for people to read articles from their own countries (Thelwall & Maflahi, 2015), although removing country self-citations makes little difference and so it is not the main cause of the difference between the Scopus citation counts and the Mendeley reader counts. All ongoing sources of bias can be partly corrected for by comparing readers to citers for older articles that have accrued a substantial fraction of their readers and citers (Figure 3). This correction method is imperfect because it does not account for unequal international changes in use of Mendeley from the base period used for normalisation (Figure 5, Figure 6) and because not all readers declare a national affiliation. Correcting for changes in the national proportions of Mendeley readers over time (Figure 10), has almost no effect, however, because the correction factors are very small.

The citation-corrected Mendeley readership count calculation $ACCFYNRC_{c,y}$ (Figure 3) or the double corrected version (Figure 10) are likely to be equivalent in practice. The advantage of the first version is its simplicity for interpretation by non-experts, whereas the more complex version may be more accurate. Nevertheless, both formulae include a number of simplifications and unproven assumptions. Arguably, however, this is not a greater problem than the assumptions that need to be made about nationally-aggregated citation-based indicators because the degree to which the citation database coverage reflects the strengths and weaknesses of a country's research can greatly affect the results in both direct and indirect ways. These direct and indirect effects of citation database coverage are impossible to accurately estimate, undermining the value of all international comparisons based on citations or readers. A corollary of these observations is that comprehensive coverage of international research rather than a sampling approach based on fields is needed to give the most credible, albeit imperfect, impact comparisons.

The above conclusions are based on two weak sources of evidence: a literature-informed rational argument about likely sources of bias in the results, and a literature-informed analysis of the results based upon a visual inspection of the graphs. Unfortunately, there is no gold standard against which to compare the different methods and so this seems to be an unavoidable limitation for any attempt to justify a new national research impact indicator. None of the methods are perfect and, in the absence of a gold standard, the choice between methods that have broadly credible calculation methods and give broadly credible results is likely to require a rational argument that relies upon a weak evidence base. Thus, any policy conclusions drawn from the outcomes should take into account the weakness of the evidence.

The results suggest, but do not prove, that Mendeley readership data can reveal trends about a year earlier than can citation counts, as evidenced by the earlier overtaking of the UK and USA by Italy and of the USA by the UK in the readership data. This evidence is complex, however, because it is not clear why the Mendeley readership indicators differ from the Scopus citation indicators, because national biases in Mendeley uptake account for only minor differences. Possible reasons for the difference between readership and citation patterns include: international biases (Thelwall & Maflahi, 2015), such as European researchers tending to cite each other; countries tending to produce research that attracts early readers and citers, so the time advantage of Mendeley favours them; countries tending to publish research that it particularly interesting to the Mendeley reader demographic, such as younger researchers. Of these, there is no evidence about whether the last two are valid and no evidence about the relative strengths of the three possible factors and there may be other more important factors. Hence, although the objective of this research was to assess whether Mendeley readership data could give more stable results than Scopus citation counts, an unexpected finding is that they give different results and the reason for the difference is unclear.

The pattern shown in apparently the most accurate graph (Figure 10), would probably not be reflected in a similar exercise for the whole of Scopus because of the differences in coverage. For example the UK and USA would presumably benefit from the inclusion of a large number of heath related articles in the rest of Scopus and their lines would be higher as a result. Nevertheless, the direction and broad nature of the increases or decreases for each country are credible because they are based on the same set of subjects, although not exactly the same set of journals. In this context, the performance of Italy is particularly impressive. It has changed from being below the world average research impact for these 26 subjects to being world leading (at least compared to the other countries analysed) between 2009 and 2013, sustaining this leading position until 2015. This performance is likely to have been influenced by the use of journal impact factors and citation counts in Italian research assessments informing funding decisions (Abramo & D'Angelo, in press). There is a weaker indirect citation-based incentive in the UK, which also increased over time, but no similar incentive in Germany, despite also experiencing a substantial increase (Figure 3). Overall, however, the results are consistent with citation-based incentives driving dramatic increases in national research impact performance.

# 8. Acknowledgement

The idea to use Mendeley readership data for early impact calculations to compare countries originated from Dr Steven Hill, Head of Research Policy at the Higher Education Funding Council for England.